\documentclass[apl,aip,amsmath,amssymb,preprint]{revtex4-2}
\usepackage{bm}% bold math
\usepackage{amssymb,amssymb,graphicx}
\usepackage{mathtools}
\usepackage{graphics}
\usepackage{booktabs}
\usepackage{xcolor}
\usepackage{dcolumn}
\usepackage{subcaption}

\newcommand{\p}{$\%$}
\newcommand{\pn}{$\mathrm{P{_{N_2}}}$}
\newcommand{\pa}{$\mathrm{P{_{Ar}}}$}

\newcommand{\Ts}{T$_\mathrm{s}$}

\newcommand{\rn}{$\mathrm{R{_{N_2}}}$}

\newcommand{\Sc}{ScN$_\mathrm{x}$}
\newcommand{\dg}{${^{\circ}}$}

\begin{document}

\title{Stranski-Krastanov Growth of Disordered \Sc~Thin Films on MgO(100): Influence of Defect Densities on Electronic Structure and Transport Properties}
%\tnotetext[mytitlenote]{ \href{http://www.ctan.org/tex-archive/macros/latex/contrib/elsarticle}{CTAN}.}


\author{Susmita Chowdhury}
\email{susmita.chowdhury@kemi.uu.se} 
\affiliation{Applied Science Department, Institute of Engineering and Technology, DAVV, Indore, 452 017, India}
\affiliation{Inorganic Chemistry, Department of Chemistry - {\AA}ngstr\"{o}m Laboratory, Uppsala University, Box 538, SE-751 21 Uppsala, Sweden}
\author{Rachana Gupta}
\affiliation{Applied Science Department, Institute of Engineering and Technology, DAVV, Indore, 452 017, India}
\author{Najnin Bano}
\affiliation{UGC-DAE Consortium for Scientific Research, University Campus, Khandwa Road, Indore, 452 001, India}
\author{Yogesh Kumar}
\affiliation{UGC-DAE Consortium for Scientific Research, University Campus, Khandwa Road, Indore, 452 001, India}
\affiliation{Department of Physics, Multanimal Modi College, Modinagar 201204, India}
\author{Shashi Prakash}
\affiliation{Applied Science Department, Institute of Engineering and Technology, DAVV, Indore, 452 017, India}
\author{Dinesh Kumar Shukla}
\affiliation{UGC-DAE Consortium for Scientific Research, University Campus, Khandwa Road, Indore, 452 001, India}
\author{Vasant G. Sathe}
\affiliation{UGC-DAE Consortium for Scientific Research, University Campus, Khandwa Road, Indore, 452 001, India}
\author{Mukul Gupta}
\email{mgupta@csr.res.in} 
\affiliation{UGC-DAE Consortium for Scientific Research, University Campus, Khandwa Road, Indore, 452 001, India}

%\date{\today}

\begin{abstract}
We report a nascent real time Stranski-Krastanov growth of reactively sputtered {\Sc} thin films on MgO(100). The epitaxial growth was limited to 5\,nm at a substrate temperature ({\Ts}) of $ \sim$ 25\dg{C} while the self sustaining epitaxial nature along the [100] azimuth was retained up to 25\,nm in {\Ts} = 250 and 500\dg{C} samples due to enhanced adatom mobility. At {\Ts} = 700{\dg}C, the film showed half order in-situ RHEED pattern, with forbidden (\textit{hkl}) planes indicating N deficient hcp Sc-N phase. Presence of defect densities \textit{i.e.}, N vacancies and O interstitials leads to a disorder in {\Sc} system with weak localization effect and appearance of Raman relaxed first order transverse and longitudinal optical phonon modes and further leads to metal like Seebeck coefficient. Higher grain boundaries at {\Ts} = 25{\dg}C and higher N out-diffusion at {\Ts} = 700{\dg}C paves way for incorporation of higher oxygen interstitial in these samples. 
\end{abstract}

\maketitle

%\section{Introduction}\label{Introduction}
Scandium nitride (ScN) is an emerging early semiconducting transition metal nitride, and is a sole stable candidate within group 3 nitrides, exhibiting an indirect band gap of 0.9\,eV.~\cite{biswas2019development,al2004surface} ScN is an attractive thermoelectric material with a typical power factor of 2.5\,-3.5\,$\times$\,10$^{-3}$ W\,m$^{-1}$\,K$^{-2}$ and a thermal conductivity of 10-12\,W\,m$^{-1}$\,K$^{-1}$. ~\cite{burmistrova2013thermoelectric,kerdsongpanya2011anomalously,kerdsongpanya2017phonon} Furthermore, it serves as a template for a defect-free growth of GaN and acts as a substitute for InN to serve as a heterostructure in the ScN / GaN system.~\cite{haseman2020cathodoluminescence} ScN has recently been demonstrated as a flexible optoelectronic material~\cite{mukhopadhyay2024flexible} while suitable alloying has led Sc$_{1-x}$Al$_{x}$N to be an excellent piezoelectric material with a piezoelectric coefficient of 27.6 pC\,N$^{-1}$.~\cite{akiyama2009enhancement}

Albeit, ScN has widely been deposited at high substrate temperatures ({\Ts\,$\geq$}\,550{\dg}C) using molecular beam epitaxy, sputtering \textit{etc.}, we have established that room-temperature sputter deposited polycrystalline ScN thin films show comparable optical and mechanical properties to epitaxial ScN thin films.~\cite{chowdhury2022detailed} Point defects in ScN affects the structural and electronic properties which in turn affects the application based perspectives.~\cite{kerdsongpanya2012effect,kumagai2018point} ScN is known to crystallize in a cubic rock salt-type B1 structure, but distortion in the local octahedral symmetry has been observed,~\cite{chowdhury2022detailed,chowdhury2024electronic} which can be attributed either to finite incorporation of oxygen due to low heat of formation of Sc-O bonds and/or presence of N vacancies.~\cite{mukhopadhyay2024flexible,chowdhury2024electronic} In addition, commonly observed nitrogen vacancies in ScN creates defect states in the valence band near to the Fermi level enhancing the thermoelectric power factor.~\cite{kerdsongpanya2012effect}

Two key questions remain unexplored in the ScN system, (i) If epitaxial growth of ScN thin films is possible at lower substrate temperature ({\Ts} $\leq$ 550{\dg}C) on single crystalline substrate motivated by the 12$^{th}$ UN sustainable goal: responsible consumption and production, and (ii) If N out-diffusion takes place in ScN at higher {\Ts} and leads to N vacancies? For a widely studied TiN counterpart, the grain boundary N-out diffusion has been demonstrated at an annealing temperature as low as 250{\dg}C and a volume type diffusion has been reported at 500{\dg}C~.\cite{chowdhury2021role} Thus, it makes the case study for ScN system which needs to be investigated. 

To explore, we adopted a dcMS set up equipped with \textit{in situ} RHEED facility to study the growth evolution of {\Sc} thin film at different {\Ts}$\approx$~25, 250, 500 and 700{\dg}C and probe the correlated electronic, transport and thermoelectric properties. Our work highlights, for the first time through in-situ RHEED, possibility of epitaxial growth of sputter deposited {\Sc} thin films on MgO(100) up to a critical thickness and confirms N out diffusion at {\Ts} = 700{\dg}C leads to N depletion and paves way for incorporation of O interstitial point defects.

%However, only a handful of studies have adopted the surface sensitive \textit{in-situ} reflection high energy electron diffraction (RHEED) analysis to gain insight into the real-time growth behavior of samples~\cite{al2000molecular,casamento2019molecular}. So far, to our knowledge, no report is available on the \textit{in-situ} growth of reactively sputtered ScN thin films combining RHEED analysis. We adopted a direct current magnetron sputtering (dcMS) equipped with \textit{in situ} RHEED facility to study {\Sc} thin film samples. The findings of the present work impart knowledge on the growth evolution of {\Sc} thin film at different {\Ts}. 

%\section{Methodology}\label{METHODOLOGY}

Single crystalline MgO (100) substrates were pre-heated at 500{\dg}C for 1\,hour and were allowed to cool down to room temperature. The differentially pumped \textit{in-situ} kSA instruments real-time RHEED acquisition system was equipped with a Staib electron gun with an operational acceleration voltage of 35\,kV. More details of the set-up can be found elsewhere.~\cite{gupta2021situ} The RHEED patterns of the MgO substrates were recorded followed by growth evolution of {\Sc} thin film samples when deposited using dcMS technique at {\Ts} = RT ($ \approx$~25\,$^{\circ}$C), 250, 500 and 700{\dg}C. The partial pressures of Ar (\pa) and N$ _{2} $ (\pn) flow of 37.5 and 12.5\,sccm were kept constant for all the adopted {\Ts}, resulting in a relative N$ _{2} $ flow [{\rn} = {\pn}/({\pa} + {\pn})] of 25{\p}. Complementary out-of-plane x-ray diffraction (XRD) measurements were carried out using Bruker XRD system using Cu-K$ \alpha $ x-rays in Bragg-Brentano geometry. The acceleration voltage and current were set at 40\,kV and 40\,mA, respectively.

The electronic structure of {\Sc} thin film samples were probed using soft x-ray absorption (SXAS) technique around Sc \textit{L$ _\mathrm{3,2} $}, N and O~\textit{K}-edges at the synchrotron facility BL-01, Indus-2, RRCAT, Indore~\cite{phase2014development} in total electron yield mode. The temperature dependent resistivity on all samples were performed in a physical property measurement system (PPMS). The sample deposited at {\Ts} = 700{\dg}C exhibited high resistance, rendering it unmeasurable. The thermopower measurements were performed via pulse method~\cite{ahad2019setup} conducted on two samples deposited at {\Ts} = 250 and 500{\dg}C. Due to high resistance of the other two samples ({\Ts} = RT and 700{\dg}C), they were deemed unsuitable for thermopower measurements. The temperature dependent Raman measurements were recorded by Horiba JY HR-800 spectrometer equipped with a He-Ne source (633\,nm) and the set up was calibrated using the peak position of single crystalline Si(111) substrate. 

%\section{Results and Discussion}

%\subsection{\textit{In-situ} surface sensitive structural properties of {\Sc} thin films}\label{XRD}

\begin{figure*} [!h]
	\begin{center}
    \vspace{-1 mm}
	\includegraphics 
	[width=0.45\textwidth] {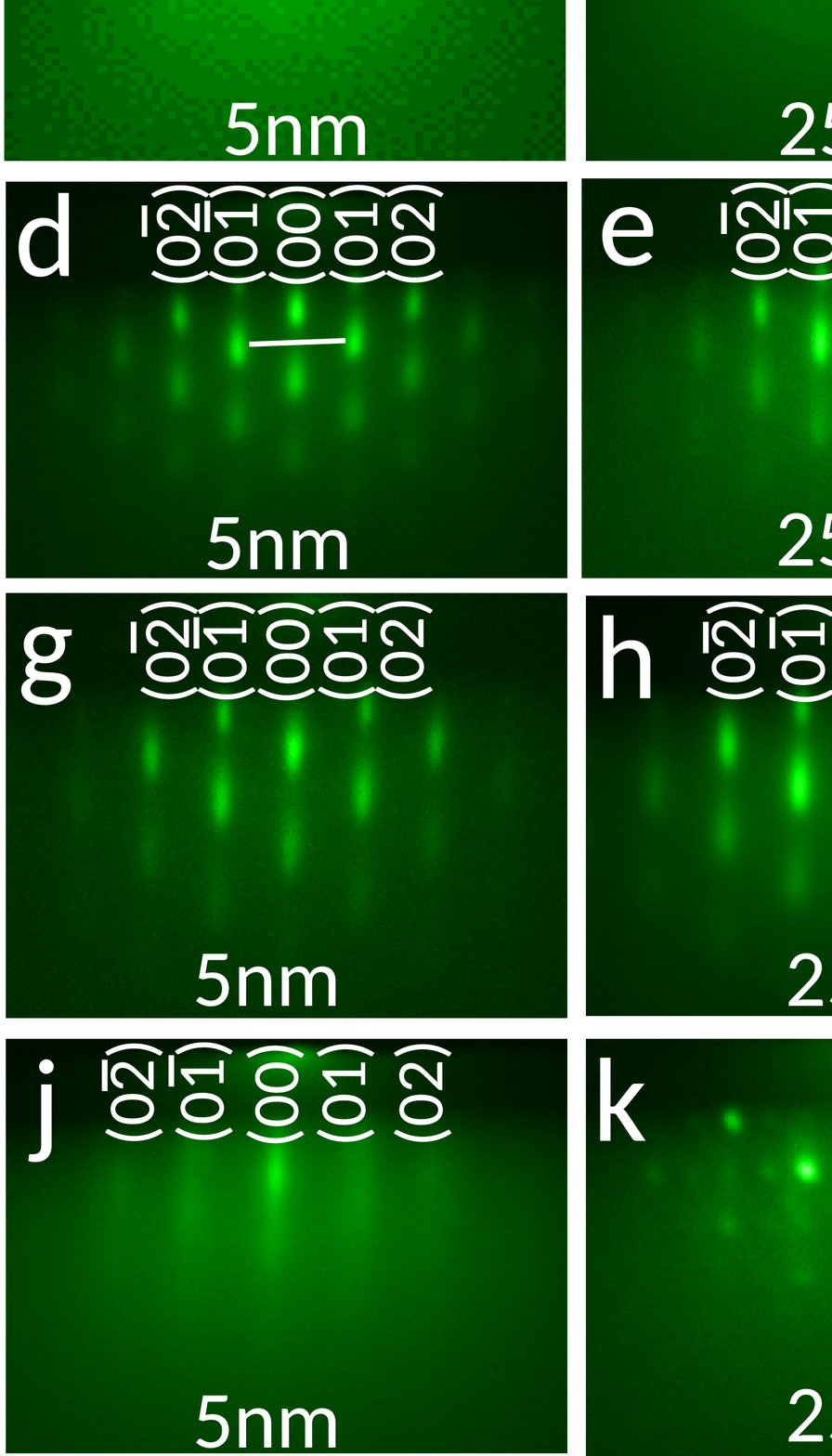}
    \includegraphics 
    [width=0.28\textwidth] {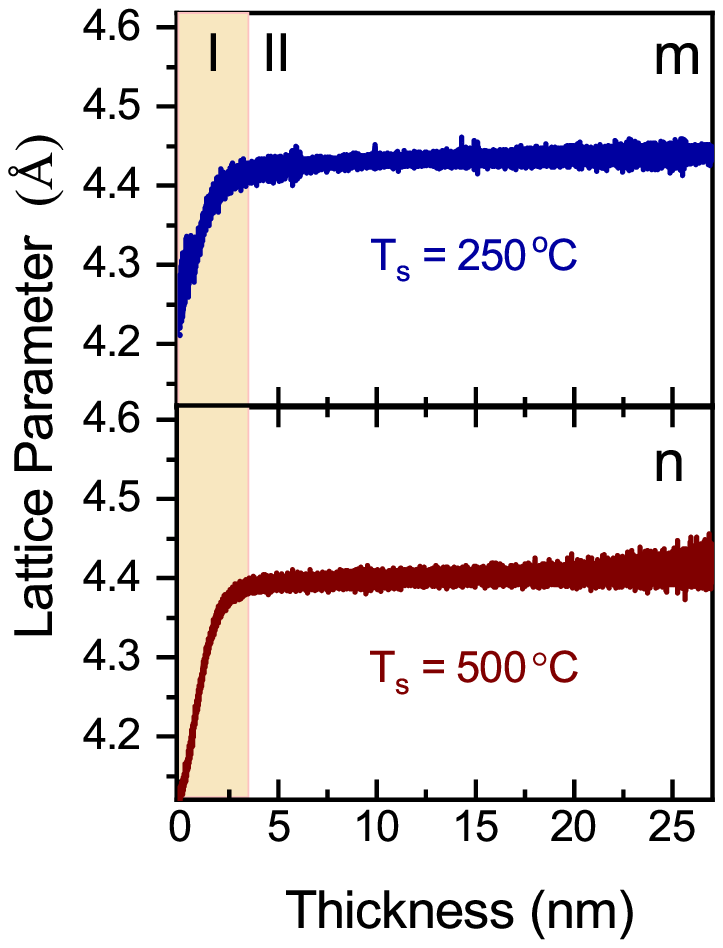}
	\vspace{-1 mm}
	\caption{RHEED images of {\Sc} thin films deposited on MgO (100) substrate at {\Ts} = 25$ ^{\circ} $C (a, b, c); 250$ ^{\circ} $C (d, e, f); $500 ^{\circ} $C (g, h, i) and 700$ ^{\circ} $C (j, k, l). Real time variation of lattice parameter of {\Sc} thin films deposited on MgO (100) substrate at {\Ts} = 250$ ^{\circ} $C (m) and 500$ ^{\circ} $C (n) analyzed through \textit{in-situ} RHEED.}
	\label{rheed}
	\end{center}
\end{figure*}

Prior to deposition, the bare MgO substrates showed sharp streaky pattern with Kikuchi lines along the [100] azimuth, indicating enhanced crystalline quality with low roughness (not shown) post annealing. Once, the deposition was initiated, fig.~\ref{rheed}(a)-(l) demonstrates the \textit{in-situ} RHEED pattern of {\Sc} thin film samples with indexed (\textit{hk}) planes deposited at various {\Ts}. At {\Ts} = 25{\dg}C, as can be seen from fig.~\ref{rheed}(a)-(c), initially the {\Sc} thin film sample follows the substrate induced epitaxy up to 5\,nm, but with further increase in thickness, a clear evolution from spotty to distinct ring-like patterns can be seen, which signifies transformation from rough 2D to polycrystalline 3D growth of the sample. The random distribution of grains is due to the highest surface energy of ScN (111) than ScN (100) plane, which in turn leads to preferable grain growth of (111) grains at lower adatom mobility. Moreover, the atomic misfit of $ \sim $7.3\% between MgO (100) and ScN (100) is expected to induce interfacial strain hindering the epitaxial growth at RT and allows for strain relaxation with further increase in thickness. The appearance of broken rings suggests preferential texturing of the sample. 

In contrast, for the samples deposited at higher {\Ts}~=~250 and 500{\dg}C [fig.~\ref{rheed}(d)-(f) and fig.~\ref{rheed}(g)-(i)], streaky pattern appears at the initial stage of growth and sustained up to 25\,nm but as the thickness increases an admixture of streaky and faint ring-like features emerges that can be seen more clearly at the film thickness of 50\,nm. This suggests a smooth 2D like growth as opposed to RT samples, due to enhanced adatom mobility leading to preferential epitaxial island ordering up to higher thickness regime ($ \leq $~25\,nm). However, surpassing a critical thickness again leads to random nucleation of the {\Sc} crystals with mixed type growth, as the surface barrier height is still high in comparison to the combined thermal energy and substrate mediated strain energy effect. This suggests Stranski-Krastanov growth of the {\Sc} samples. The real time variation of in-plane lattice parameter (LP) obtained from the (0$\bar{1}$) and (01) spacing (shown by horizontal line in fig.~\ref{rheed}(d)) for {\Sc} thin films deposited at {\Ts} = 250 and 500{\dg}C are shown in fig.~\ref{rheed}(m) and (n). Here, region I depicts the strain relaxation zone of the {\Sc} samples. Due to higher LP of ScN compared to MgO (4.21\,{\AA}), typically up to 3.4\,nm of thickness, both films sustain strain and with further increase in the thickness, the relaxation of interfacial strain is observed as constant LP in region II upto 30\,nm where the epitaxial growth has been observed. The relaxed LP obtained for 250{\dg}C and 500{\dg}C deposited {\Sc} samples are 4.43 and 4.39\,{\AA}, respectively and a lower value of LP at {\Ts} = 500{\dg}C could be due to low N content in the sample.
      
In case of the sample deposited at {\Ts} = 700{\dg}C, non-faceted 2D like growth can be observed during the initial deposition up to 5\,nm and later on, evolution of half order features with fundamental intense spots are noticeable demonstrating growth of flat surfaces.~\cite{park2002c} Since the RHEED patterns differ appreciably from the previously observed features at lower {\Ts}, there is a certain possibility of deviation of the crystal structure from the rocksalt type ScN, due to N out-diffusion at high {\Ts}. Hence, out-of-plane XRD can provide better insight on the crystal structure of {\Sc} samples.

%\subsection{Bulk sensitive structural properties of {\Sc} thin films}\label{XRD} 

\begin{figure} 
\vspace{-5 mm}
	\begin{center}
	\includegraphics 
	[width=0.3\textwidth] {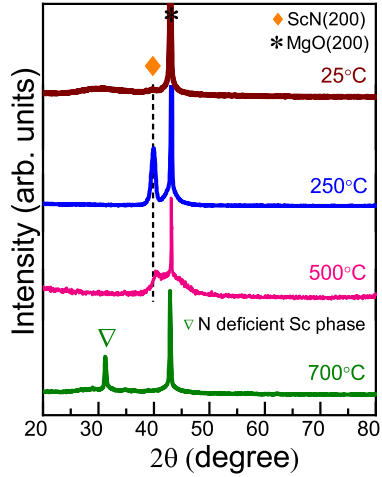}
	\vspace{0 mm}
	\caption{XRD pattern of ScN thin film samples deposited at {\Ts} = 25, 250, 500, and 700$ ^{\circ} $C.}
	\label{xrdrheed}
	\end{center}
\end{figure}

Figure~\ref{xrdrheed} shows the XRD pattern of {\Sc} thin film samples deposited at various {\Ts}. The 25{\dg}C deposited sample showed a small peak of (200) reflection plane at around 39.9$ ^{\circ} $ yielding an out-of-plane LP of 4.515\,\AA. The appearance of a broad hump at $ \sim $25-35$ ^{\circ} $ is suspected to be from the clay used to mount the sample. %Thus the mixed orientations demonstrate polycrystalline nature of the B1 NaCl type rocksalt crystal structure of RT deposited sample in line with our \textit{in-situ} RHEED results. 
For {\Ts} = 250{\dg}C sample, the crystalline quality of the sample increases with an intense (200) peak and the obtained LP was the same as in the RT sample. With further increase in {\Ts} to 500{\dg}C, the only visible difference was appearance of less prominent (200) peak which shifts to a slightly higher diffraction angle compared to the {\Ts} = 250{\dg}C sample. Thus, the out-of-plane LP contracts for this sample compared to RT and {\Ts} = 250{\dg}C sample. Thus, it further confirms the cubic rocksalt crystal structure of the samples and unlike RHEED, we did not observe any additional reflection planes along the out-of-plane direction. But, with further increase in {\Ts} to 700{\dg}C, appearance of an intense peak at $ \sim $31.2$ ^{\circ} $ was observed with absence of the (200) peak. The peak position indicates forbidden (\textit{hkl}) plane and a clear deviation from the cubic rocksalt structure. We attribute it to some hexagonal close packed N-deficient Sc-N phase, which occurs due to N out-diffusion at this growth temeperature.  

%\subsection{Electronic structure of {\Sc} thin films}\label{SXAS}

\begin{figure*}
	\begin{center}
	\includegraphics 
	[width=0.7\textwidth] {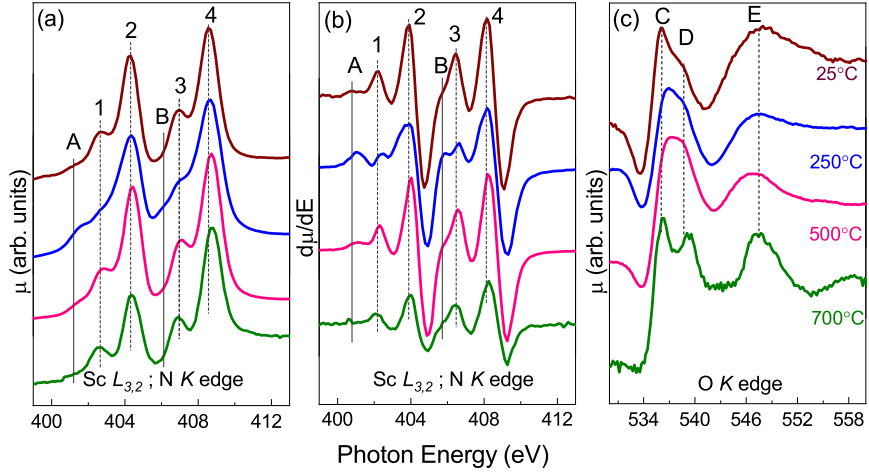}
	\vspace{0 mm}
	\caption{XAS spectra of {\Sc} thin film samples deposited at {\Ts} = 25, 250, 500, and 700$ ^{\circ} $C taken at Sc $L_{3,2} $ and N K-edge (a) its derivative (b) and that of O $K$-edge (c).}
	\label{xasrheed}
	\end{center}
\end{figure*}

Figure~\ref{xasrheed}(a) shows the SXAS spectra of {\Sc} thin film samples deposited at different {\Ts} around Sc \textit{L$ _\mathrm{3,2} $} and N \textit{K}-edge. The N \textit{K}-edge appears very close to Sc \textit{L$ _\mathrm{3} $}-edge and moreover Sc is prone to oxidation often leading to additional crystal field splitting.~\cite{chowdhury2022detailed} To understand it more qualitatively fig.~\ref{xasrheed}(b) shows the first order derivative of fig.~\ref{xasrheed}(a) with respect to the photon energy. For all the {\Sc} samples, four prominent features `1', `2', `3' and `4' appear at around 402.6, 404.3, 406.9 and 408.6\,eV, while features `A' and `B' appear at around 401.4 and 406\,eV. 
\par
The first four features (1 to 4) likely occur due to the spin split of Sc 3d states into \textit{L$ _{3} $} and \textit{L$ _{2} $} and further hybridization of the O 2p states with the Sc 3d states result in crystal field splitting into t$ _\mathrm{2g} $ ($ \pi $-bonds) and e$ _\mathrm{g} $ ($ \sigma $-bonds) orbitals. The features `A' and `B' are resultant of Sc 3d-N 2p hybridizations resulting in the crystal field split and resembles similarity to earlier reports.~\cite{chowdhury2022detailed} Among the samples, the visible difference is observed in the relative intensity ratio (Intensity$_{A+B}$/Intensity$_{1+3}$) of the features `A' and `B' to `1' and `3' and is highest for 250{\dg}C deposited sample. For 700{\dg}C deposited sample, the intensity of features `A' and `B' nearly diminishes and looks similar to Sc \textit{L}-edge Sc$ _{2} $O$ _{3} $ spectrum~\cite{de19902} with crystal field splitting (10Dq) of 1.7($ \pm $0.3)\,eV (1.8\,eV reported for Sc$ _{2} $O$ _{3} $). However, any definite conclusions could not be drawn only by looking at the 10Dq values around Sc \textit{L}-edge as this edge is known to be affected by multiplet effects.~\cite{de19902,chen1997nexafs} Thus, it can be concluded that for 250{\dg}C sample, possibly the nitrogen (oxygen) content is highest (lowest) while it is lowest (highest) for 700{\dg}C sample. However, the real picture can be confirmed by probing the O \textit{K}-edge of the samples. 
\par
Figure~\ref{xasrheed}(c) shows O \textit{K}-edge spectra of samples. The doublet features appear at around 536~-~539\,eV labeled as `C' and `D' while the higher energy feature `E' peaks at around 546{--}548\,eV. The doublet feature appears due to the hybridization of O 2p$\pi$+Sc3d and O 2p$\sigma$+Sc3d leading to crystal field splitting into t$ _\mathrm{2g} $ and e$ _\mathrm{g} $ states. Feature `E' appears due to the hybridization of O 2p orbitals with higher energy Sc 4sp orbitals. The signature of the doublet features is relatively less pronounced for 250 and 500{\dg}C samples, and  as compared to 700{\dg}C samples. For the 700{\dg}C sample, a clear splitting around the doublet appears and the spectrum resembles previously reported O \textit{K}-edge of Sc$ _{2} $O$ _{3} $.~\cite{de1989oxygen} 10Dq = 3($ \pm $0.3)\,eV was obtained for this sample which falls in the range of 10Dq = 3.3\,eV reported for Sc$ _{2} $O$ _{3} $~\cite{de1989oxygen} considering the energy resolution. Thus, the observations confirm the presence of highest oxygen incorporation in the samples deposited at 700{\dg}C. Although the {\rn} flow is constant for all the samples, the possible reason for oxygen incorporation is essentially N out-diffusion at this high temperature. This leads to N vacancies in the anion site providing the route for oxygen accommodation in interstitials. Usually samples with high amount of oxygen should have higher resistivity compared to pure ScN samples. This paves the need to perform electrical measurements which is also required to measure the thermoelectric responses of the samples.

%\subsection{Electrical and thermoelectric properties of {\Sc} thin films}\label{Seebeck}

\begin{figure*}
	\begin{center}
		\includegraphics 
		[width=0.5\textwidth] {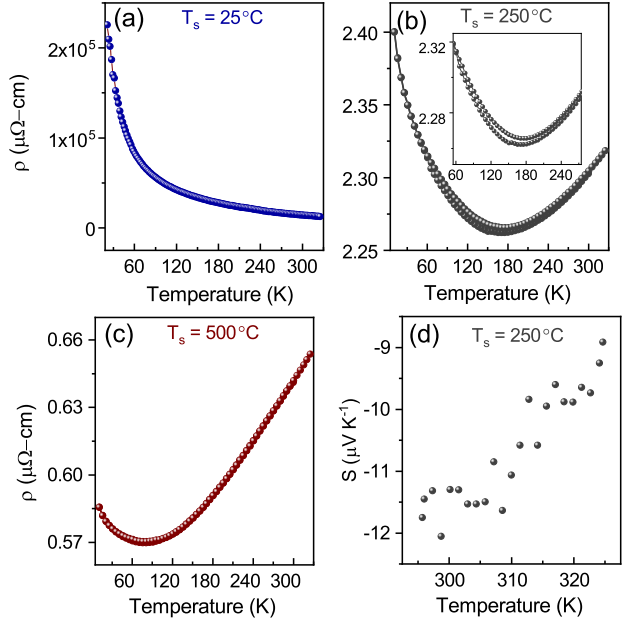}
		\vspace{0 mm}
		\caption{Temperature dependent resistivity of {\Sc} thin film samples deposited at {\Ts} = 25$ ^{\circ} $C (a), 250$ ^{\circ} $C (b) and 500$ ^{\circ} $C (c). The temperature dependent Seebeck co-efficient for {\Ts} = 250$ ^{\circ} $C (d) is also shown. {\Ts} = 700$ ^{\circ} $C sample exhibited high resistance rendering it to be unmeasurable. Due to the high resistance of {\Ts} = 25$ ^{\circ} $C and 700$ ^{\circ} $C samples, they were deemed unsuitable for thermopower measurements.}
		\label{seebeckrheed}
	\end{center}
\end{figure*}

Figure~\ref{seebeckrheed}(a), (b) and (c) shows the temperature dependent resistivity of 25, 250 and 500{\dg}C deposited {\Sc} samples. For 25{\dg}C sample, the resistivity gradually decreases from 2.25 to 0.13\,$\mu\Omega$-cm with increase in temperature from 21 to 325\,K, and shows a semiconducting behavior.~\cite{gall1998microstructure} However, extensively an opposite trend exhibiting positive slope (typical metallic nature) have been reported for ScN bulk~\cite{al2020properties} and thin films~\cite{cetnar2018electronic,rao2020high} in earlier reports. Due to low enthalpy of formation of Sc-O, even presence of slight oxygen drives the ScN system to possess metal like conductivity. In the present study, we attribute this semiconducting nature to the presence of higher amount of oxygen as defects rather than assigning it to the formation of pure ScN. For 250 and 500{\dg}C deposited samples, an interesting behavior is noted. Although the resistivity values of the samples purely lie in the metallic region for these two samples, initially resistivity gradually decreases with increase in temperature and reaching a minima at a critical temperature an upturn is observed up to 325\,K. The resistivity minima is observed at around 168\,K for the {\Ts} = 250{\dg}C deposited sample and at around 68\,K for {\Ts} = 500{\dg}C deposited sample. The resistivity upturn deviating from metallic behavior at low temperatures is a signature of weak localization effect due to presence of defect densities and reflects disordered {\Sc} system.~\cite{bergmann1984weak} An ambiguous hysteresis loop is noted for the first time in the ScN system during the heating and cooling cycle as shown in the inset of fig.~\ref{seebeckrheed}(b) as a magnified view. The hysteresis effect is observed in the temperature region of 62-304\,K. Usually, such hysteresis in the resistivity data is indicative of a first order phase transition involving structural transitions. But further analysis such as temperature dependent Raman or XRD measurements are required to probe the real phenomena behind such observations. Apart from Sc-N bonds, simultaneous bond formation of Sc to oxygen was also confirmed from our soft XAS study. This possibly gives rise to such an interesting trend in 250 and 500{\dg}C samples. 

The temperature dependent Seebeck data are shown in fig.~\ref{seebeckrheed}(d) for the 250{\dg}C samples. Presence of relatively high defect densities significantly lower than previously reported Seebeck values in similar temperature ranges, suggesting suppressed thermoelectric performance in the present case.~\cite{biswas2019development,le2018effect} For the 250{\dg}C sample, the Seebeck coefficient increases in magnitude from approximately -12 to -9\,$ \mu $V K$ ^{-1} $ with increase in temperature from 295 to 325\,K. For 500{\dg}C sample, these values are even lower (not shown). Usually such low Seebeck values are typical for metals.~\cite{macia2015thermoelectric}    

%\subsection{Vibrational Raman modes of {\Sc} thin films}\label{Raman}

\begin{figure*} 
	\begin{center}
		\includegraphics 
		[width=0.6\textwidth] {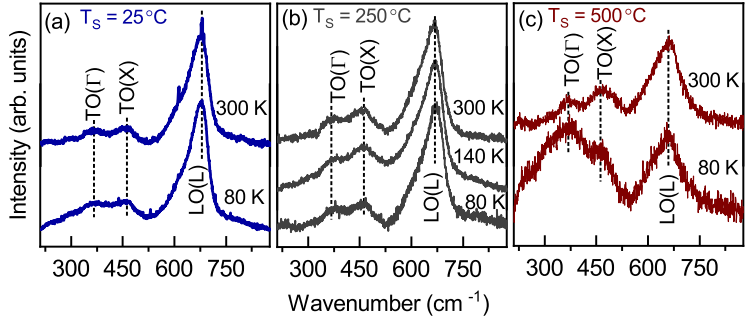}
		\vspace{0 mm}
		\caption{Temperature dependent Raman of {\Sc} thin film samples deposited at {\Ts} = 25$ ^{\circ} $C (a), 250$ ^{\circ} $C (b) and 500$ ^{\circ} $C. For {\Ts} = 25$ ^{\circ} $C and 500$ ^{\circ} $C {\Sc} samples, the Raman data was measured at 80 and 300\,K, while for {\Ts} = 500$ ^{\circ} $C, the Raman data was measured at 80, 140 and 300\,K corroborating to the three different region observed from transport measurement.}
		\label{ramanrheed}
	\end{center}
\end{figure*}

Fig.~\ref{ramanrheed}(a), (b) and (c) shows the temperature dependent vibrational spectra of 25, 250 and 500$ ^{\circ} $C deposited {\Sc} samples. Although first order Raman modes are forbidden in cubic rocksalt structures,~\cite{burstein1965selection} appearance of transverse optical (TO: from $\Gamma$ and X symmetry point in the Brillouin zone) and longitudinal optical (LO: from the L symmetry point)~\cite{grumbel2024band,travaglini1986electronic,gall2001vibrational} phonon mode confirms Raman relaxation and indicates distortion from local cubic symmetry due to presence of point defects for all the {\Sc} samples throughout the temperature range. Here, the broad TO modes appear at TO($\Gamma$):\,$\approx$\,365 and TO(X):\,$\approx$\,462\,cm$ ^{-1} $, while the intense LO(L) appears at $\approx$\,677 ($\pm$\,2)\,cm$ ^{-1} $ for {\Ts} = 25$ ^{\circ} $C, $\approx$\,668 ($\pm$\,2)\,cm$ ^{-1} $ at 80 and 140\,K and $\approx$\,665 ($\pm$\,2)\,cm$ ^{-1} $ at 300\,K for {\Ts} = 250{\dg}C and $\approx$\,661 ($\pm$\,2)\,cm$ ^{-1} $ for {\Ts} = 500{\dg}C, respectively. The gradual phonon softening from {\Ts} = 25 to 500{\dg}C sample accompanied by broadened phonon modes at 500{\dg}C is indicative of highest defect density present in this sample. The absence of any additional features for {\Ts} = 250{\dg}C sample within the Raman temperature range negates the presence of structural transition around the hysteresis region observed from our transport measurement in this sample. However, local distortions can be present within the sample which will explored in future.    

%\section*{Conclusion}

\textit{In-situ} real time RHEED analysis during sputtering suggest Stranski-Krastanov growth of {\Sc} thin films on MgO(100) substrates. The lower adatom mobility at RT promotes epitaxial growth only up to 5\,nm and lattice mismatch induced strain assists polycrystalline growth beyond it. Increase in the {\Ts} enhances the epitaxial island ordering up to $ \approx $25~nm, but later on evolves to random grain growth. However, at {\Ts} = 700{\dg}C, half order features in RHEED  and forbidden (\textit{hkl}) plane in the out-of-plane XRD suggests evolution of N deficient hcp Sc-N phase. Soft XAS study reveals lowest N content for the {\Ts} = 700{\dg}C paving way for incorporation of oxygen interstitials rendering to measure the electrical resistivity exceeding the limit. The RT deposited sample showed semiconducting behavior due to presence higher oxygen incorporation due to higher grain boundaries while the resistivity values for 250 and 500{\dg}C deposited samples lie in the metallic region. This also results in metal-like Seebeck coefficient values for these samples with weak localization effect. The presence of defect densities limit the thermoelectric performance and further distort the octahedral symmetry of {\Sc} thin films.

%\section*{Acknowledgments}

We are grateful to UGC-DAE CSR, Indore for providing financial support through CRS project. Thanks are due to A. Rathore, L. Behera, R. Sah, A Wadikar and S. Kalal for the assistance provided in experiments. SC thanks S. Jamal and H. K. Singh for fruitful discussions.

\section*{}

%\bibliography{bibliography}

\begin{thebibliography}{31}%
	\makeatletter
	\providecommand \@ifxundefined [1]{%
		\@ifx{#1\undefined}
	}%
	\providecommand \@ifnum [1]{%
		\ifnum #1\expandafter \@firstoftwo
		\else \expandafter \@secondoftwo
		\fi
	}%
	\providecommand \@ifx [1]{%
		\ifx #1\expandafter \@firstoftwo
		\else \expandafter \@secondoftwo
		\fi
	}%
	\providecommand \natexlab [1]{#1}%
	\providecommand \enquote  [1]{``#1''}%
	\providecommand \bibnamefont  [1]{#1}%
	\providecommand \bibfnamefont [1]{#1}%
	\providecommand \citenamefont [1]{#1}%
	\providecommand \href@noop [0]{\@secondoftwo}%
	\providecommand \href [0]{\begingroup \@sanitize@url \@href}%
	\providecommand \@href[1]{\@@startlink{#1}\@@href}%
	\providecommand \@@href[1]{\endgroup#1\@@endlink}%
	\providecommand \@sanitize@url [0]{\catcode `\\12\catcode `\$12\catcode
		`\&12\catcode `\#12\catcode `\^12\catcode `\_12\catcode `\%12\relax}%
	\providecommand \@@startlink[1]{}%
	\providecommand \@@endlink[0]{}%
	\providecommand \url  [0]{\begingroup\@sanitize@url \@url }%
	\providecommand \@url [1]{\endgroup\@href {#1}{\urlprefix }}%
	\providecommand \urlprefix  [0]{URL }%
	\providecommand \Eprint [0]{\href }%
	\providecommand \doibase [0]{https://doi.org/}%
	\providecommand \selectlanguage [0]{\@gobble}%
	\providecommand \bibinfo  [0]{\@secondoftwo}%
	\providecommand \bibfield  [0]{\@secondoftwo}%
	\providecommand \translation [1]{[#1]}%
	\providecommand \BibitemOpen [0]{}%
	\providecommand \bibitemStop [0]{}%
	\providecommand \bibitemNoStop [0]{.\EOS\space}%
	\providecommand \EOS [0]{\spacefactor3000\relax}%
	\providecommand \BibitemShut  [1]{\csname bibitem#1\endcsname}%
	\let\auto@bib@innerbib\@empty
	%</preamble>
	\bibitem [{\citenamefont {Biswas}\ and\ \citenamefont
		{Saha}(2019)}]{biswas2019development}%
	\BibitemOpen
	\bibfield  {author} {\bibinfo {author} {\bibfnamefont {B.}~\bibnamefont
			{Biswas}}\ and\ \bibinfo {author} {\bibfnamefont {B.}~\bibnamefont {Saha}},\
	}\href@noop {} {\bibfield  {journal} {\bibinfo  {journal} {Physical Review
				Materials}\ }\textbf {\bibinfo {volume} {3}},\ \bibinfo {pages} {020301}
		(\bibinfo {year} {2019})}\BibitemShut {NoStop}%
	\bibitem [{\citenamefont {Al-Brithen}, \citenamefont {Smith},\ and\
		\citenamefont {Gall}(2004)}]{al2004surface}%
	\BibitemOpen
	\bibfield  {author} {\bibinfo {author} {\bibfnamefont {H.~A.}\ \bibnamefont
			{Al-Brithen}}, \bibinfo {author} {\bibfnamefont {A.~R.}\ \bibnamefont
			{Smith}},\ and\ \bibinfo {author} {\bibfnamefont {D.}~\bibnamefont {Gall}},\
	}\href@noop {} {\bibfield  {journal} {\bibinfo  {journal} {Physical Review
				B}\ }\textbf {\bibinfo {volume} {70}},\ \bibinfo {pages} {045303} (\bibinfo
		{year} {2004})}\BibitemShut {NoStop}%
	\bibitem [{\citenamefont {Burmistrova}\ \emph {et~al.}(2013)\citenamefont
		{Burmistrova}, \citenamefont {Maassen}, \citenamefont {Favaloro},
		\citenamefont {Saha}, \citenamefont {Salamat}, \citenamefont {Rui~Koh},
		\citenamefont {Lundstrom}, \citenamefont {Shakouri},\ and\ \citenamefont
		{Sands}}]{burmistrova2013thermoelectric}%
	\BibitemOpen
	\bibfield  {author} {\bibinfo {author} {\bibfnamefont {P.~V.}\ \bibnamefont
			{Burmistrova}}, \bibinfo {author} {\bibfnamefont {J.}~\bibnamefont
			{Maassen}}, \bibinfo {author} {\bibfnamefont {T.}~\bibnamefont {Favaloro}},
		\bibinfo {author} {\bibfnamefont {B.}~\bibnamefont {Saha}}, \bibinfo {author}
		{\bibfnamefont {S.}~\bibnamefont {Salamat}}, \bibinfo {author} {\bibfnamefont
			{Y.}~\bibnamefont {Rui~Koh}}, \bibinfo {author} {\bibfnamefont {M.~S.}\
			\bibnamefont {Lundstrom}}, \bibinfo {author} {\bibfnamefont {A.}~\bibnamefont
			{Shakouri}},\ and\ \bibinfo {author} {\bibfnamefont {T.~D.}\ \bibnamefont
			{Sands}},\ }\href@noop {} {\bibfield  {journal} {\bibinfo  {journal} {Journal
				of Applied Physics}\ }\textbf {\bibinfo {volume} {113}},\ \bibinfo {pages}
		{153704} (\bibinfo {year} {2013})}\BibitemShut {NoStop}%
	\bibitem [{\citenamefont {Kerdsongpanya}\ \emph {et~al.}(2011)\citenamefont
		{Kerdsongpanya}, \citenamefont {Van~Nong}, \citenamefont {Pryds},
		\citenamefont {{\v{Z}}ukauskait{\.e}}, \citenamefont {Jensen}, \citenamefont
		{Birch}, \citenamefont {Lu}, \citenamefont {Hultman}, \citenamefont
		{Wingqvist},\ and\ \citenamefont {Eklund}}]{kerdsongpanya2011anomalously}%
	\BibitemOpen
	\bibfield  {author} {\bibinfo {author} {\bibfnamefont {S.}~\bibnamefont
			{Kerdsongpanya}}, \bibinfo {author} {\bibfnamefont {N.}~\bibnamefont
			{Van~Nong}}, \bibinfo {author} {\bibfnamefont {N.}~\bibnamefont {Pryds}},
		\bibinfo {author} {\bibfnamefont {A.}~\bibnamefont {{\v{Z}}ukauskait{\.e}}},
		\bibinfo {author} {\bibfnamefont {J.}~\bibnamefont {Jensen}}, \bibinfo
		{author} {\bibfnamefont {J.}~\bibnamefont {Birch}}, \bibinfo {author}
		{\bibfnamefont {J.}~\bibnamefont {Lu}}, \bibinfo {author} {\bibfnamefont
			{L.}~\bibnamefont {Hultman}}, \bibinfo {author} {\bibfnamefont
			{G.}~\bibnamefont {Wingqvist}},\ and\ \bibinfo {author} {\bibfnamefont
			{P.}~\bibnamefont {Eklund}},\ }\href@noop {} {\bibfield  {journal} {\bibinfo
			{journal} {Applied Physics Letters}\ }\textbf {\bibinfo {volume} {99}},\
		\bibinfo {pages} {232113} (\bibinfo {year} {2011})}\BibitemShut {NoStop}%
	\bibitem [{\citenamefont {Kerdsongpanya}\ \emph {et~al.}(2017)\citenamefont
		{Kerdsongpanya}, \citenamefont {Hellman}, \citenamefont {Sun}, \citenamefont
		{Koh}, \citenamefont {Lu}, \citenamefont {Van~Nong}, \citenamefont {Simak},
		\citenamefont {Alling},\ and\ \citenamefont
		{Eklund}}]{kerdsongpanya2017phonon}%
	\BibitemOpen
	\bibfield  {author} {\bibinfo {author} {\bibfnamefont {S.}~\bibnamefont
			{Kerdsongpanya}}, \bibinfo {author} {\bibfnamefont {O.}~\bibnamefont
			{Hellman}}, \bibinfo {author} {\bibfnamefont {B.}~\bibnamefont {Sun}},
		\bibinfo {author} {\bibfnamefont {Y.~K.}\ \bibnamefont {Koh}}, \bibinfo
		{author} {\bibfnamefont {J.}~\bibnamefont {Lu}}, \bibinfo {author}
		{\bibfnamefont {N.}~\bibnamefont {Van~Nong}}, \bibinfo {author}
		{\bibfnamefont {S.~I.}\ \bibnamefont {Simak}}, \bibinfo {author}
		{\bibfnamefont {B.}~\bibnamefont {Alling}},\ and\ \bibinfo {author}
		{\bibfnamefont {P.}~\bibnamefont {Eklund}},\ }\href@noop {} {\bibfield
		{journal} {\bibinfo  {journal} {Physical Review B}\ }\textbf {\bibinfo
			{volume} {96}},\ \bibinfo {pages} {195417} (\bibinfo {year}
		{2017})}\BibitemShut {NoStop}%
	\bibitem [{\citenamefont {Haseman}\ \emph {et~al.}(2020)\citenamefont
		{Haseman}, \citenamefont {Noesges}, \citenamefont {Shields}, \citenamefont
		{Cetnar}, \citenamefont {Reed}, \citenamefont {Al-Atabi}, \citenamefont
		{Edgar},\ and\ \citenamefont {Brillson}}]{haseman2020cathodoluminescence}%
	\BibitemOpen
	\bibfield  {author} {\bibinfo {author} {\bibfnamefont {M.~S.}\ \bibnamefont
			{Haseman}}, \bibinfo {author} {\bibfnamefont {B.~A.}\ \bibnamefont
			{Noesges}}, \bibinfo {author} {\bibfnamefont {S.}~\bibnamefont {Shields}},
		\bibinfo {author} {\bibfnamefont {J.~S.}\ \bibnamefont {Cetnar}}, \bibinfo
		{author} {\bibfnamefont {A.~N.}\ \bibnamefont {Reed}}, \bibinfo {author}
		{\bibfnamefont {H.~A.}\ \bibnamefont {Al-Atabi}}, \bibinfo {author}
		{\bibfnamefont {J.~H.}\ \bibnamefont {Edgar}},\ and\ \bibinfo {author}
		{\bibfnamefont {L.~J.}\ \bibnamefont {Brillson}},\ }\href@noop {} {\bibfield
		{journal} {\bibinfo  {journal} {APL Materials}\ }\textbf {\bibinfo {volume}
			{8}},\ \bibinfo {pages} {081103} (\bibinfo {year} {2020})}\BibitemShut
	{NoStop}%
	\bibitem [{\citenamefont {Mukhopadhyay}\ \emph {et~al.}(2024)\citenamefont
		{Mukhopadhyay}, \citenamefont {Rao}, \citenamefont {Rawat}, \citenamefont
		{Pillai}, \citenamefont {Garbrecht},\ and\ \citenamefont
		{Saha}}]{mukhopadhyay2024flexible}%
	\BibitemOpen
	\bibfield  {author} {\bibinfo {author} {\bibfnamefont {D.}~\bibnamefont
			{Mukhopadhyay}}, \bibinfo {author} {\bibfnamefont {D.}~\bibnamefont {Rao}},
		\bibinfo {author} {\bibfnamefont {R.~S.}\ \bibnamefont {Rawat}}, \bibinfo
		{author} {\bibfnamefont {A.~I.~K.}\ \bibnamefont {Pillai}}, \bibinfo {author}
		{\bibfnamefont {M.}~\bibnamefont {Garbrecht}},\ and\ \bibinfo {author}
		{\bibfnamefont {B.}~\bibnamefont {Saha}},\ }\href@noop {} {\bibfield
		{journal} {\bibinfo  {journal} {Nano Letters}\ }\textbf {\bibinfo {volume}
			{24}},\ \bibinfo {pages} {14493} (\bibinfo {year} {2024})}\BibitemShut
	{NoStop}%
	\bibitem [{\citenamefont {Akiyama}\ \emph {et~al.}(2009)\citenamefont
		{Akiyama}, \citenamefont {Kamohara}, \citenamefont {Kano}, \citenamefont
		{Teshigahara}, \citenamefont {Takeuchi},\ and\ \citenamefont
		{Kawahara}}]{akiyama2009enhancement}%
	\BibitemOpen
	\bibfield  {author} {\bibinfo {author} {\bibfnamefont {M.}~\bibnamefont
			{Akiyama}}, \bibinfo {author} {\bibfnamefont {T.}~\bibnamefont {Kamohara}},
		\bibinfo {author} {\bibfnamefont {K.}~\bibnamefont {Kano}}, \bibinfo {author}
		{\bibfnamefont {A.}~\bibnamefont {Teshigahara}}, \bibinfo {author}
		{\bibfnamefont {Y.}~\bibnamefont {Takeuchi}},\ and\ \bibinfo {author}
		{\bibfnamefont {N.}~\bibnamefont {Kawahara}},\ }\href@noop {} {\bibfield
		{journal} {\bibinfo  {journal} {Advanced Materials}\ }\textbf {\bibinfo
			{volume} {21}},\ \bibinfo {pages} {593} (\bibinfo {year} {2009})}\BibitemShut
	{NoStop}%
	\bibitem [{\citenamefont {Chowdhury}\ \emph {et~al.}(2022)\citenamefont
		{Chowdhury}, \citenamefont {Gupta}, \citenamefont {Rajput}, \citenamefont
		{Tayal}, \citenamefont {Rao}, \citenamefont {Sekhar}, \citenamefont
		{Prakash}, \citenamefont {Rajagopalan}, \citenamefont {Jha}, \citenamefont
		{Saha} \emph {et~al.}}]{chowdhury2022detailed}%
	\BibitemOpen
	\bibfield  {author} {\bibinfo {author} {\bibfnamefont {S.}~\bibnamefont
			{Chowdhury}}, \bibinfo {author} {\bibfnamefont {R.}~\bibnamefont {Gupta}},
		\bibinfo {author} {\bibfnamefont {P.}~\bibnamefont {Rajput}}, \bibinfo
		{author} {\bibfnamefont {A.}~\bibnamefont {Tayal}}, \bibinfo {author}
		{\bibfnamefont {D.}~\bibnamefont {Rao}}, \bibinfo {author} {\bibfnamefont
			{R.}~\bibnamefont {Sekhar}}, \bibinfo {author} {\bibfnamefont
			{S.}~\bibnamefont {Prakash}}, \bibinfo {author} {\bibfnamefont
			{R.}~\bibnamefont {Rajagopalan}}, \bibinfo {author} {\bibfnamefont
			{S.}~\bibnamefont {Jha}}, \bibinfo {author} {\bibfnamefont {B.}~\bibnamefont
			{Saha}}, \emph {et~al.},\ }\href@noop {} {\bibfield  {journal} {\bibinfo
			{journal} {Materialia}\ }\textbf {\bibinfo {volume} {22}},\ \bibinfo {pages}
		{101375} (\bibinfo {year} {2022})}\BibitemShut {NoStop}%
	\bibitem [{\citenamefont {Kerdsongpanya}, \citenamefont {Alling},\ and\
		\citenamefont {Eklund}(2012)}]{kerdsongpanya2012effect}%
	\BibitemOpen
	\bibfield  {author} {\bibinfo {author} {\bibfnamefont {S.}~\bibnamefont
			{Kerdsongpanya}}, \bibinfo {author} {\bibfnamefont {B.}~\bibnamefont
			{Alling}},\ and\ \bibinfo {author} {\bibfnamefont {P.}~\bibnamefont
			{Eklund}},\ }\href@noop {} {\bibfield  {journal} {\bibinfo  {journal}
			{Physical Review B—Condensed Matter and Materials Physics}\ }\textbf
		{\bibinfo {volume} {86}},\ \bibinfo {pages} {195140} (\bibinfo {year}
		{2012})}\BibitemShut {NoStop}%
	\bibitem [{\citenamefont {Kumagai}, \citenamefont {Tsunoda},\ and\
		\citenamefont {Oba}(2018)}]{kumagai2018point}%
	\BibitemOpen
	\bibfield  {author} {\bibinfo {author} {\bibfnamefont {Y.}~\bibnamefont
			{Kumagai}}, \bibinfo {author} {\bibfnamefont {N.}~\bibnamefont {Tsunoda}},\
		and\ \bibinfo {author} {\bibfnamefont {F.}~\bibnamefont {Oba}},\ }\href@noop
	{} {\bibfield  {journal} {\bibinfo  {journal} {Physical Review Applied}\
		}\textbf {\bibinfo {volume} {9}},\ \bibinfo {pages} {034019} (\bibinfo {year}
		{2018})}\BibitemShut {NoStop}%
	\bibitem [{\citenamefont {Chowdhury}\ \emph {et~al.}(2024)\citenamefont
		{Chowdhury}, \citenamefont {Singh}, \citenamefont {Honnali}, \citenamefont
		{Greczynski}, \citenamefont {Eklund}, \citenamefont {le~Febvrier},\ and\
		\citenamefont {Magnuson}}]{chowdhury2024electronic}%
	\BibitemOpen
	\bibfield  {author} {\bibinfo {author} {\bibfnamefont {S.}~\bibnamefont
			{Chowdhury}}, \bibinfo {author} {\bibfnamefont {N.~K.}\ \bibnamefont
			{Singh}}, \bibinfo {author} {\bibfnamefont {S.~K.}\ \bibnamefont {Honnali}},
		\bibinfo {author} {\bibfnamefont {G.}~\bibnamefont {Greczynski}}, \bibinfo
		{author} {\bibfnamefont {P.}~\bibnamefont {Eklund}}, \bibinfo {author}
		{\bibfnamefont {A.}~\bibnamefont {le~Febvrier}},\ and\ \bibinfo {author}
		{\bibfnamefont {M.}~\bibnamefont {Magnuson}},\ }\href@noop {} {\bibfield
		{journal} {\bibinfo  {journal} {Physical Review B}\ }\textbf {\bibinfo
			{volume} {110}},\ \bibinfo {pages} {115139} (\bibinfo {year}
		{2024})}\BibitemShut {NoStop}%
	\bibitem [{\citenamefont {Chowdhury}\ \emph {et~al.}(2021)\citenamefont
		{Chowdhury}, \citenamefont {Gupta}, \citenamefont {Prakash}, \citenamefont
		{Behera},\ and\ \citenamefont {Gupta}}]{chowdhury2021role}%
	\BibitemOpen
	\bibfield  {author} {\bibinfo {author} {\bibfnamefont {S.}~\bibnamefont
			{Chowdhury}}, \bibinfo {author} {\bibfnamefont {R.}~\bibnamefont {Gupta}},
		\bibinfo {author} {\bibfnamefont {S.}~\bibnamefont {Prakash}}, \bibinfo
		{author} {\bibfnamefont {L.}~\bibnamefont {Behera}},\ and\ \bibinfo {author}
		{\bibfnamefont {M.}~\bibnamefont {Gupta}},\ }\href@noop {} {\bibfield
		{journal} {\bibinfo  {journal} {Applied Surface Science}\ }\textbf {\bibinfo
			{volume} {564}},\ \bibinfo {pages} {150430} (\bibinfo {year}
		{2021})}\BibitemShut {NoStop}%
	\bibitem [{\citenamefont {Gupta}\ \emph {et~al.}(2021)\citenamefont {Gupta}
		\emph {et~al.}}]{gupta2021situ}%
	\BibitemOpen
	\bibfield  {author} {\bibinfo {author} {\bibfnamefont {M.}~\bibnamefont
			{Gupta}} \emph {et~al.},\ }\href@noop {} {\bibfield  {journal} {\bibinfo
			{journal} {Journal of Crystal Growth}\ }\textbf {\bibinfo {volume} {560}},\
		\bibinfo {pages} {126049} (\bibinfo {year} {2021})}\BibitemShut {NoStop}%
	\bibitem [{\citenamefont {Phase}\ \emph {et~al.}(2014)\citenamefont {Phase},
		\citenamefont {Gupta}, \citenamefont {Potdar}, \citenamefont {Behera},
		\citenamefont {Sah},\ and\ \citenamefont {Gupta}}]{phase2014development}%
	\BibitemOpen
	\bibfield  {author} {\bibinfo {author} {\bibfnamefont {D.}~\bibnamefont
			{Phase}}, \bibinfo {author} {\bibfnamefont {M.}~\bibnamefont {Gupta}},
		\bibinfo {author} {\bibfnamefont {S.}~\bibnamefont {Potdar}}, \bibinfo
		{author} {\bibfnamefont {L.}~\bibnamefont {Behera}}, \bibinfo {author}
		{\bibfnamefont {R.}~\bibnamefont {Sah}},\ and\ \bibinfo {author}
		{\bibfnamefont {A.}~\bibnamefont {Gupta}},\ }in\ \href@noop {} {\emph
		{\bibinfo {booktitle} {AIP Conference Proceedings}}},\ Vol.\ \bibinfo
	{volume} {1591}\ (\bibinfo {organization} {American Institute of Physics},\
	\bibinfo {year} {2014})\ pp.\ \bibinfo {pages} {685--686}\BibitemShut
	{NoStop}%
	\bibitem [{\citenamefont {Ahad}\ and\ \citenamefont
		{Shukla}(2019)}]{ahad2019setup}%
	\BibitemOpen
	\bibfield  {author} {\bibinfo {author} {\bibfnamefont {A.}~\bibnamefont
			{Ahad}}\ and\ \bibinfo {author} {\bibfnamefont {D.}~\bibnamefont {Shukla}},\
	}\href@noop {} {\bibfield  {journal} {\bibinfo  {journal} {Review of
				Scientific Instruments}\ }\textbf {\bibinfo {volume} {90}} (\bibinfo {year}
		{2019})}\BibitemShut {NoStop}%
	\bibitem [{\citenamefont {Park}\ \emph {et~al.}(2002)\citenamefont {Park},
		\citenamefont {D’Arcy-Gall}, \citenamefont {Gall}, \citenamefont {Kim},
		\citenamefont {Desjardins},\ and\ \citenamefont {Greene}}]{park2002c}%
	\BibitemOpen
	\bibfield  {author} {\bibinfo {author} {\bibfnamefont {S.}~\bibnamefont
			{Park}}, \bibinfo {author} {\bibfnamefont {J.}~\bibnamefont {D’Arcy-Gall}},
		\bibinfo {author} {\bibfnamefont {D.}~\bibnamefont {Gall}}, \bibinfo {author}
		{\bibfnamefont {Y.-W.}\ \bibnamefont {Kim}}, \bibinfo {author} {\bibfnamefont
			{P.}~\bibnamefont {Desjardins}},\ and\ \bibinfo {author} {\bibfnamefont
			{J.}~\bibnamefont {Greene}},\ }\href@noop {} {\bibfield  {journal} {\bibinfo
			{journal} {Journal of applied physics}\ }\textbf {\bibinfo {volume} {91}},\
		\bibinfo {pages} {3644} (\bibinfo {year} {2002})}\BibitemShut {NoStop}%
	\bibitem [{\citenamefont {De~Groot}\ \emph {et~al.}(1990)\citenamefont
		{De~Groot}, \citenamefont {Fuggle}, \citenamefont {Thole},\ and\
		\citenamefont {Sawatzky}}]{de19902}%
	\BibitemOpen
	\bibfield  {author} {\bibinfo {author} {\bibfnamefont {F.}~\bibnamefont
			{De~Groot}}, \bibinfo {author} {\bibfnamefont {J.}~\bibnamefont {Fuggle}},
		\bibinfo {author} {\bibfnamefont {B.}~\bibnamefont {Thole}},\ and\ \bibinfo
		{author} {\bibfnamefont {G.}~\bibnamefont {Sawatzky}},\ }\href@noop {}
	{\bibfield  {journal} {\bibinfo  {journal} {Physical Review B}\ }\textbf
		{\bibinfo {volume} {41}},\ \bibinfo {pages} {928} (\bibinfo {year}
		{1990})}\BibitemShut {NoStop}%
	\bibitem [{\citenamefont {Chen}(1997)}]{chen1997nexafs}%
	\BibitemOpen
	\bibfield  {author} {\bibinfo {author} {\bibfnamefont {J.~G.}\ \bibnamefont
			{Chen}},\ }\href@noop {} {\bibfield  {journal} {\bibinfo  {journal} {Surface
				Science Reports}\ }\textbf {\bibinfo {volume} {30}},\ \bibinfo {pages} {1}
		(\bibinfo {year} {1997})}\BibitemShut {NoStop}%
	\bibitem [{\citenamefont {de~Groot}\ \emph {et~al.}(1990)\citenamefont
		{de~Groot}, \citenamefont {Fuggle}, \citenamefont {Thole},\ and\
		\citenamefont {Sawatzky}}]{de1989oxygen}%
	\BibitemOpen
	\bibfield  {author} {\bibinfo {author} {\bibfnamefont {F.~M.~F.}\
			\bibnamefont {de~Groot}}, \bibinfo {author} {\bibfnamefont {J.~C.}\
			\bibnamefont {Fuggle}}, \bibinfo {author} {\bibfnamefont {B.~T.}\
			\bibnamefont {Thole}},\ and\ \bibinfo {author} {\bibfnamefont {G.~A.}\
			\bibnamefont {Sawatzky}},\ }\href {https://doi.org/10.1103/PhysRevB.41.928}
	{\bibfield  {journal} {\bibinfo  {journal} {Phys. Rev. B}\ }\textbf {\bibinfo
			{volume} {41}},\ \bibinfo {pages} {928} (\bibinfo {year} {1990})}\BibitemShut
	{NoStop}%
	\bibitem [{\citenamefont {Gall}\ \emph {et~al.}(1998)\citenamefont {Gall},
		\citenamefont {Petrov}, \citenamefont {Madsen}, \citenamefont {Sundgren},\
		and\ \citenamefont {Greene}}]{gall1998microstructure}%
	\BibitemOpen
	\bibfield  {author} {\bibinfo {author} {\bibfnamefont {D.}~\bibnamefont
			{Gall}}, \bibinfo {author} {\bibfnamefont {I.}~\bibnamefont {Petrov}},
		\bibinfo {author} {\bibfnamefont {L.}~\bibnamefont {Madsen}}, \bibinfo
		{author} {\bibfnamefont {J.-E.}\ \bibnamefont {Sundgren}},\ and\ \bibinfo
		{author} {\bibfnamefont {J.}~\bibnamefont {Greene}},\ }\href@noop {}
	{\bibfield  {journal} {\bibinfo  {journal} {Journal of Vacuum Science \&
				Technology A: Vacuum, Surfaces, and Films}\ }\textbf {\bibinfo {volume}
			{16}},\ \bibinfo {pages} {2411} (\bibinfo {year} {1998})}\BibitemShut
	{NoStop}%
	\bibitem [{\citenamefont {Al-Atabi}\ \emph {et~al.}(2020)\citenamefont
		{Al-Atabi}, \citenamefont {Zheng}, \citenamefont {Cetnar}, \citenamefont
		{Look}, \citenamefont {Cahill},\ and\ \citenamefont
		{Edgar}}]{al2020properties}%
	\BibitemOpen
	\bibfield  {author} {\bibinfo {author} {\bibfnamefont {H.}~\bibnamefont
			{Al-Atabi}}, \bibinfo {author} {\bibfnamefont {Q.}~\bibnamefont {Zheng}},
		\bibinfo {author} {\bibfnamefont {J.~S.}\ \bibnamefont {Cetnar}}, \bibinfo
		{author} {\bibfnamefont {D.}~\bibnamefont {Look}}, \bibinfo {author}
		{\bibfnamefont {D.~G.}\ \bibnamefont {Cahill}},\ and\ \bibinfo {author}
		{\bibfnamefont {J.~H.}\ \bibnamefont {Edgar}},\ }\href@noop {} {\bibfield
		{journal} {\bibinfo  {journal} {Applied Physics Letters}\ }\textbf {\bibinfo
			{volume} {116}} (\bibinfo {year} {2020})}\BibitemShut {NoStop}%
	\bibitem [{\citenamefont {Cetnar}\ \emph {et~al.}(2018)\citenamefont {Cetnar},
		\citenamefont {Reed}, \citenamefont {Badescu}, \citenamefont {Vangala},
		\citenamefont {Smith},\ and\ \citenamefont {Look}}]{cetnar2018electronic}%
	\BibitemOpen
	\bibfield  {author} {\bibinfo {author} {\bibfnamefont {J.~S.}\ \bibnamefont
			{Cetnar}}, \bibinfo {author} {\bibfnamefont {A.~N.}\ \bibnamefont {Reed}},
		\bibinfo {author} {\bibfnamefont {S.~C.}\ \bibnamefont {Badescu}}, \bibinfo
		{author} {\bibfnamefont {S.}~\bibnamefont {Vangala}}, \bibinfo {author}
		{\bibfnamefont {H.~A.}\ \bibnamefont {Smith}},\ and\ \bibinfo {author}
		{\bibfnamefont {D.~C.}\ \bibnamefont {Look}},\ }\href@noop {} {\bibfield
		{journal} {\bibinfo  {journal} {Applied Physics Letters}\ }\textbf {\bibinfo
			{volume} {113}},\ \bibinfo {pages} {192104} (\bibinfo {year}
		{2018})}\BibitemShut {NoStop}%
	\bibitem [{\citenamefont {Rao}\ \emph {et~al.}(2020)\citenamefont {Rao},
		\citenamefont {Biswas}, \citenamefont {Flores}, \citenamefont {Chatterjee},
		\citenamefont {Garbrecht}, \citenamefont {Koh}, \citenamefont {Bhatia},
		\citenamefont {Pillai}, \citenamefont {Hopkins}, \citenamefont
		{Martin-Gonzalez} \emph {et~al.}}]{rao2020high}%
	\BibitemOpen
	\bibfield  {author} {\bibinfo {author} {\bibfnamefont {D.}~\bibnamefont
			{Rao}}, \bibinfo {author} {\bibfnamefont {B.}~\bibnamefont {Biswas}},
		\bibinfo {author} {\bibfnamefont {E.}~\bibnamefont {Flores}}, \bibinfo
		{author} {\bibfnamefont {A.}~\bibnamefont {Chatterjee}}, \bibinfo {author}
		{\bibfnamefont {M.}~\bibnamefont {Garbrecht}}, \bibinfo {author}
		{\bibfnamefont {Y.~R.}\ \bibnamefont {Koh}}, \bibinfo {author} {\bibfnamefont
			{V.}~\bibnamefont {Bhatia}}, \bibinfo {author} {\bibfnamefont {A.~I.~K.}\
			\bibnamefont {Pillai}}, \bibinfo {author} {\bibfnamefont {P.~E.}\
			\bibnamefont {Hopkins}}, \bibinfo {author} {\bibfnamefont {M.}~\bibnamefont
			{Martin-Gonzalez}}, \emph {et~al.},\ }\href@noop {} {\bibfield  {journal}
		{\bibinfo  {journal} {Applied Physics Letters}\ }\textbf {\bibinfo {volume}
			{116}},\ \bibinfo {pages} {152103} (\bibinfo {year} {2020})}\BibitemShut
	{NoStop}%
	\bibitem [{\citenamefont {Bergmann}(1984)}]{bergmann1984weak}%
	\BibitemOpen
	\bibfield  {author} {\bibinfo {author} {\bibfnamefont {G.}~\bibnamefont
			{Bergmann}},\ }\href@noop {} {\bibfield  {journal} {\bibinfo  {journal}
			{Physics Reports}\ }\textbf {\bibinfo {volume} {107}},\ \bibinfo {pages} {1}
		(\bibinfo {year} {1984})}\BibitemShut {NoStop}%
	\bibitem [{\citenamefont {Le~Febvrier}\ \emph {et~al.}(2018)\citenamefont
		{Le~Febvrier}, \citenamefont {Tureson}, \citenamefont {Stilkerich},
		\citenamefont {Greczynski},\ and\ \citenamefont {Eklund}}]{le2018effect}%
	\BibitemOpen
	\bibfield  {author} {\bibinfo {author} {\bibfnamefont {A.}~\bibnamefont
			{Le~Febvrier}}, \bibinfo {author} {\bibfnamefont {N.}~\bibnamefont
			{Tureson}}, \bibinfo {author} {\bibfnamefont {N.}~\bibnamefont {Stilkerich}},
		\bibinfo {author} {\bibfnamefont {G.}~\bibnamefont {Greczynski}},\ and\
		\bibinfo {author} {\bibfnamefont {P.}~\bibnamefont {Eklund}},\ }\href@noop {}
	{\bibfield  {journal} {\bibinfo  {journal} {Journal of Physics D: Applied
				Physics}\ }\textbf {\bibinfo {volume} {52}},\ \bibinfo {pages} {035302}
		(\bibinfo {year} {2018})}\BibitemShut {NoStop}%
	\bibitem [{\citenamefont {Maci{\'a}-Barber}(2015)}]{macia2015thermoelectric}%
	\BibitemOpen
	\bibfield  {author} {\bibinfo {author} {\bibfnamefont {E.}~\bibnamefont
			{Maci{\'a}-Barber}},\ }in\ \href@noop {} {\emph {\bibinfo {booktitle}
			{Advances and Applications}}}\ (\bibinfo  {publisher} {Taylor \& Francis
		Group, Pan Stanford},\ \bibinfo {year} {2015})\BibitemShut {NoStop}%
	\bibitem [{\citenamefont {Burstein}, \citenamefont {Johnson},\ and\
		\citenamefont {Loudon}(1965)}]{burstein1965selection}%
	\BibitemOpen
	\bibfield  {author} {\bibinfo {author} {\bibfnamefont {E.}~\bibnamefont
			{Burstein}}, \bibinfo {author} {\bibfnamefont {F.}~\bibnamefont {Johnson}},\
		and\ \bibinfo {author} {\bibfnamefont {R.}~\bibnamefont {Loudon}},\
	}\href@noop {} {\bibfield  {journal} {\bibinfo  {journal} {Physical Review}\
		}\textbf {\bibinfo {volume} {139}},\ \bibinfo {pages} {A1239} (\bibinfo
		{year} {1965})}\BibitemShut {NoStop}%
	\bibitem [{\citenamefont {Gr{\"u}mbel}\ \emph {et~al.}(2024)\citenamefont
		{Gr{\"u}mbel}, \citenamefont {Goldhahn}, \citenamefont {Feneberg},
		\citenamefont {Oshima}, \citenamefont {Dubroka},\ and\ \citenamefont
		{Ramsteiner}}]{grumbel2024band}%
	\BibitemOpen
	\bibfield  {author} {\bibinfo {author} {\bibfnamefont {J.}~\bibnamefont
			{Gr{\"u}mbel}}, \bibinfo {author} {\bibfnamefont {R.}~\bibnamefont
			{Goldhahn}}, \bibinfo {author} {\bibfnamefont {M.}~\bibnamefont {Feneberg}},
		\bibinfo {author} {\bibfnamefont {Y.}~\bibnamefont {Oshima}}, \bibinfo
		{author} {\bibfnamefont {A.}~\bibnamefont {Dubroka}},\ and\ \bibinfo {author}
		{\bibfnamefont {M.}~\bibnamefont {Ramsteiner}},\ }\href@noop {} {\bibfield
		{journal} {\bibinfo  {journal} {Physical Review Materials}\ }\textbf
		{\bibinfo {volume} {8}},\ \bibinfo {pages} {L071601} (\bibinfo {year}
		{2024})}\BibitemShut {NoStop}%
	\bibitem [{\citenamefont {Travaglini}\ \emph {et~al.}(1986)\citenamefont
		{Travaglini}, \citenamefont {Marabelli}, \citenamefont {Monnier},
		\citenamefont {Kaldis},\ and\ \citenamefont
		{Wachter}}]{travaglini1986electronic}%
	\BibitemOpen
	\bibfield  {author} {\bibinfo {author} {\bibfnamefont {G.}~\bibnamefont
			{Travaglini}}, \bibinfo {author} {\bibfnamefont {F.}~\bibnamefont
			{Marabelli}}, \bibinfo {author} {\bibfnamefont {R.}~\bibnamefont {Monnier}},
		\bibinfo {author} {\bibfnamefont {E.}~\bibnamefont {Kaldis}},\ and\ \bibinfo
		{author} {\bibfnamefont {P.}~\bibnamefont {Wachter}},\ }\href@noop {}
	{\bibfield  {journal} {\bibinfo  {journal} {Physical Review B}\ }\textbf
		{\bibinfo {volume} {34}},\ \bibinfo {pages} {3876} (\bibinfo {year}
		{1986})}\BibitemShut {NoStop}%
	\bibitem [{\citenamefont {Gall}, \citenamefont {Stoehr},\ and\ \citenamefont
		{Greene}(2001)}]{gall2001vibrational}%
	\BibitemOpen
	\bibfield  {author} {\bibinfo {author} {\bibfnamefont {D.}~\bibnamefont
			{Gall}}, \bibinfo {author} {\bibfnamefont {M.}~\bibnamefont {Stoehr}},\ and\
		\bibinfo {author} {\bibfnamefont {J.}~\bibnamefont {Greene}},\ }\href@noop {}
	{\bibfield  {journal} {\bibinfo  {journal} {Physical Review B}\ }\textbf
		{\bibinfo {volume} {64}},\ \bibinfo {pages} {174302} (\bibinfo {year}
		{2001})}\BibitemShut {NoStop}%
\end{thebibliography}

%

\end{document}